\pdfoutput=1 
\documentclass[aps,prl,twocolumn,showpacs,superscriptaddress,groupedaddress]{revtex4}  
\usepackage{graphicx}  
\usepackage{dcolumn}   
\usepackage{bm}         
\usepackage{amssymb}   
\usepackage{float}		

\begin{document}

	\widetext
%
\title{Rainbow-trapping by adiabatic tuning of intragroove plasmon coupling }
%

\author{A.O.~Montazeri} 
\affiliation{Department of Electrical \& Computer Engineering, University of Toronto, Toronto, Ontario, Canada  M5S 3G4}
\author{Y.~Fang} 
\affiliation{Department of Physics, University of Toronto, Toronto, Ontario, Canada M5S 1A4}
\author{P. Sarrafi}
\affiliation{Department of Electrical \& Computer Engineering, University of Toronto, Toronto, Ontario, Canada  M5S 3G4}
\author{N.P.~Kherani} 
\affiliation{Department of Electrical \& Computer Engineering, University of Toronto, Toronto, Ontario, Canada  M5S 3G4}
\affiliation{Department of Materials Science \& Engineering, University of Toronto, Toronto, Ontario, Canada M5S 3E4}

%
%
%
\vskip 0.25cm

\begin{abstract}
It is shown here that rendering the width of nano-grooves into a tunable parameter presents a new means of light trapping in subwavelength gratings. In gratings with groove-widths below 150 nm, the plasmon coupling between the perimeter walls of the groove becomes dominant and thus turns groove width into the principle geometric parameter in determining light trapping. Using this parameter, we investigate the prospect of tunable optical functionalities. An analytical formula is derived by treating each nano-groove as a plasmonic waveguide resonator. The resulting change in the cavity effective mode index is examined; these results are in close agreement with numerical simulations. It is shown that the tunable nano-groove resonator presented here accurately defines waveguiding, slow-light, and light-trapping regimes.
\end{abstract}
\maketitle
\begin{figure} [ht]
\includegraphics[height=55mm, width=87mm]{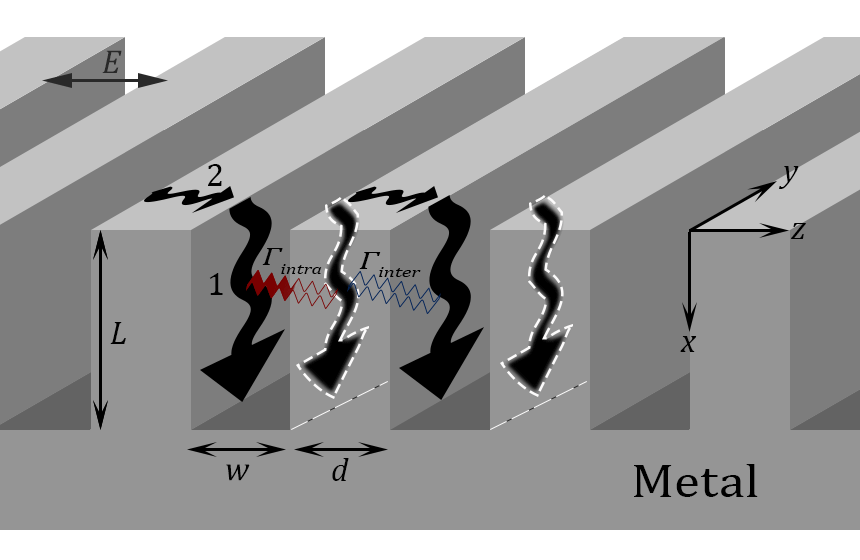}
\caption{\label{fig:Schematic} $P$-polarized radiation ($E$-field in the $z$-direction) can launch SPPs  traveling in the $x$-direction in the $xy$ plane into the grooves (labeled 1), as well as SPPs traveling in the z-direction. In narrow grooves when $w$  $\lesssim$150 nm, two Type-1 SPPs couple and form the intragroove coupling $\Gamma_{intra}$. Likewise, two resonant cavities couple when $d$ is comparable in size to the skin depth of SPPs in the metal which acts to reinforce intergroove coupling shown as $\Gamma_{inter}$. A secondary mechanism of intergroove coupling is mediated through surface SPPs traveling in the $z$-direction in the $yz$ plane (labeled 2). Only Type-1 SPPs take part in the anomalous absorption of light.}
\end{figure}

  
Surface plasmon polaritons (SPPs) have the remarkable ability to confine light into extremely subwavelength volumes such as nano-grooves \cite{Spoof_angle,Groove_confinement}. This effective field enhancement has led to their application in biology and chemistry, non-linear optics, and optoelectronics \cite{Brongersma_shalaev,Stockman_review,plasmonic_book}. When such nano-grooves are arranged in the formation of a grating, their coupling becomes possible at the right groove separation. However, each nano-groove need not be identical to its neighbors. In the presence of a gradient---or a perturbation across these weakly coupled nano-grooves---fields confined within one groove can ``flow" into the next \cite{Moire}. In this Letter, we present a new functional gradient in the structure of gratings through the tailoring of the geometry of individual nano-grooves. That is, in addition to the said inter-groove coupling, we argue that tailoring the intra-groove SPP-coupling enables an additional degree of freedom in the design and properties of subwavelength devices. In other words, nanoscale control of individual grooves in turn grants control of wave propagation over the extended surface of the grating such as that shown in Fig.~\ref{fig:Schematic}. The underlying physics of this intragroove-tailoring is shown to consist of tuning the interaction strength of SPPs on the opposing walls of each nano-groove in the strong-coupling regime \cite{Japan_nanogrooves}. Not only does this tailoring control the near-field optical properties of nano-grooves, but it also shapes the far-field optical behavior of the resulting grating through which emerge phenomena such as slow-light,``rainbow-trapping", and impedance-matching. It is this novel physical phenomena and their simple yet powerful corollaries that are discussed in this Letter.

Graded gratings are a subset of functionally-graded materials such as graded-index fiber optics \cite{graded_fiber_detailed_paper}, and more generally, structures possessing a spatial gradient in their materials composition or their geometric structure \cite{Suresh}. When there is an (effective) index gradient within an optical structure, light is naturally guided in the direction of increasing index. If the high index optical component is also a resonator (e.g. a waveguide resonator) in tune with the frequency of the external radiation---such as a nano-groove within a grating---at the given 	location, the group velocity of light significantly decreases.

We commence by studying nano-grooves as subwavelength metal-insulator-metal (MIM) waveguide resonators. The effective mode index $n_{eff}$, of such a waveguide resonator is tunable through the coupling strength of the evanescent SPP fields in the dielectric core of the groove \cite{Japan_nanogrooves}. This is done by reducing the core thickness which strengthens this intra-groove coupling $\Gamma_{intra}$. Using $\Gamma_{intra}$ (i.e. varying groove width $w$ in the right range, instead of the typical variation of groove depth or $L$  for Fabry-Perot type resonators,) enables a new plasmon-assisted design variable. When a chain of grooves with gradually decreasing groove widths is under weak inter-groove coupling, their effective mode indices gradually increase, and a functionally graded grating is achieved. Light is naturally guided in the direction of increasing $n_{eff}$, squeezed into thinner and thinner grooves, and effectively trapped at the location of the groove resonating at the frequency of impinging light.  This intra-cavity mode engineering of gratings through adjustment of $\Gamma_{intra}$ helps to clearly understand the emergence of phenomena such as slow-light and the so called ``rainbow-trapping'' effects.
 
%

Various approaches have been taken to analyze the structure of gratings and their interactions with light \cite{pendry_hole,vidal_grating,science_nanosmooth_patterning,gradient_metamat,Duke_dots}. Here, we take each groove as a metal-insulator-metal (MIM) waveguide. The odd-modes of such a waveguide have no cut-off even in the extremely narrow groove widths of less than 10 nm \cite{plasmonic_book,Perchec}. We show that the strong coupling of SPPs within the MIM structure of each groove (in the $x$-direction as shown in Fig.~\ref{fig:Schematic}) enables control over the propagation constant of SPPs over the resulting ensemble of many grooves viz., the grating itself ($z$-direction in Fig.~\ref{fig:Schematic}\textbf{}).

When a grating such as the structure shown in Fig.~\ref{fig:Schematic} is illuminated by $p$-polarized radiation ($E$-field in the $z$-direction as shown in Fig.~\ref{fig:Schematic}), SPPs are launched downwards into the grooves (labeled Type-1 SPPs) \cite{Alex_optex}. Additionally,  the horizontal flat surface of the metal between grooves, can also support SPP formation (intergroove SPPs of Type-$2$). When $w$$\lesssim$150 nm, two down-traveling SPPs couple by the virtue of their overlapping fields (intragroove coupling labelled $\Gamma_{intra}$ and illustrated as a squiggly line). Likewise,  coupling of SPPs takes place through the metal walls when the thickness $d$ of the metallic wall is comparable to the evanescent field in the metal (shown as intergroove coupling $\Gamma_{inter}$ and illustrated as a squiggly line through the metal wall). When the intergroove separation $d$ is larger than the evanescent field of the SPP in the metal, there is effectively no direct coupling through the metallic walls, and $\Gamma_{inter}$ is mediated externally through Type-$2$ surface SPPs \cite{Alex_optex}. 

While intergroove coupling deals with the transfer of energy between grooves, the intragroove coupling $\Gamma_{intra}$ acts to modify the optical cavity modes and the effective mode index  $n_{eff}$ of the groove.

\begin{figure}
\includegraphics[height=55mm, width=87mm]{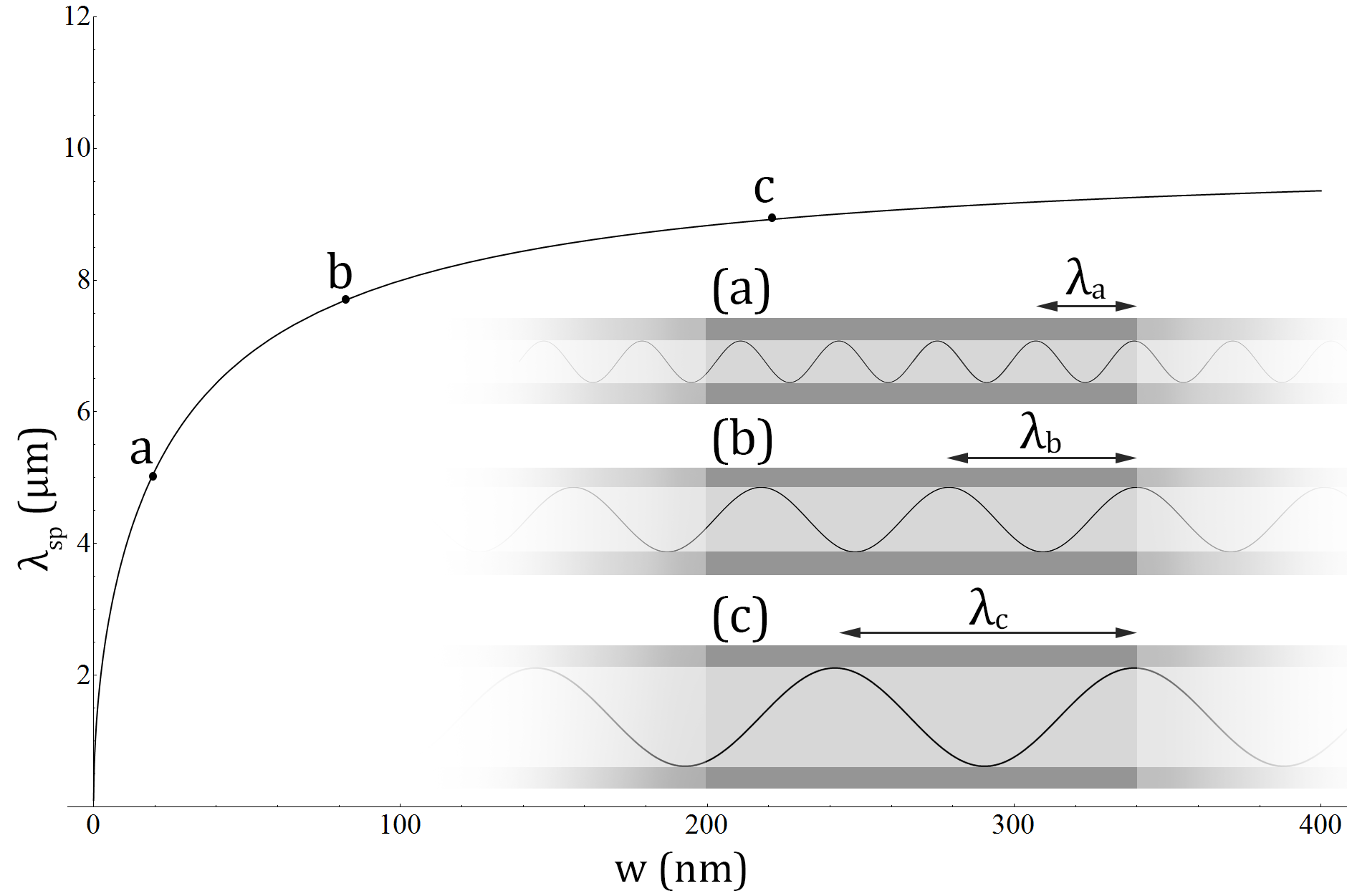}
\caption{\label{fig:open_MIM} This plot shows $\lambda_{sp}$ in an unbounded MIM waveguide as a function of dielectric core thickness $w$ with permittivity $\epsilon_1$, and a metal cladding with $\epsilon_2$. The Fabry-Perot cavity wavelength is 10 $\mu$m. $\lambda_{sp}$ at point (a) is compressed to about half this cavity wavelength due to the intragroove coupling effects. In larger gap sizes ($>$200 nm) $\lambda_{sp}$ approaches the cavity wavelength.}
\end{figure}

For an infinitely long MIM waveguide, the effects of changing the insulator (core) thickness $w$ are shown in Fig.~\ref{fig:open_MIM}. For $w$ much larger than the field confinement in metal, i.e. beyond point $c$ in Fig.~\ref{fig:open_MIM}, the surface plasmon wavelength $\lambda_{sp}$, supported by the MIM waveguide, becomes relatively insensitive to the variations in $w$. In this region $\Gamma_{intra}$ is negligible. As $w$ becomes smaller, however, the effect of intragroove coupling $\Gamma_{intra}$ on $\lambda_{sp}$ intensifies and $w$ emerges as an unfrozen degree of freedom which modifies the cavity resonant modes. As a result, resonant modes of MIM-based cavities in this regime take on a $w$-dependence in addition to $L$; whereas typical cavity resonators are only $L$-sensitive. The appearance of this effect, namely $w$-sensitivity, aids in the inception of a new tunable parameter discussed in this Letter. By smoothly tapering $\Gamma_{intra}$ through gradual and adiabatic adjustments in $w$, we introduce a gradient of $n_{eff}$ across a grating, which in turn enables control over propagation and trapping of radiation.

Further, since the other coupling process namely $\Gamma_{inter}$ is assumed to be negligibly weak here, the far-field response of a grating as a whole is determined fundamentally by the geometry of  a single unit: the groove.

It has been shown that strong light absorption by subwavelength gratings is assisted by the excitation of Type-1 SPPs within the grooves \cite{Perchec}. When all the grooves are identical and at resonance with the external radiation, the entire grating becomes absorbing at a single frequency. Yet, when there is a gradation in groove resonance across the grating, only a small number of grooves that are at or near resonance strongly absorb. The rest of the off-resonance grooves guide the wave on top and manipulate the phase of the wavefront. That is, for a broadband source of light, only spatial regions of the grating where grooves are at resonance function as a trap, and a ``rainbow trapping'' effect is observed \cite{atwater_rainbow}. 

\begin{figure} [ht]
\includegraphics[height=53mm, width=87mm]{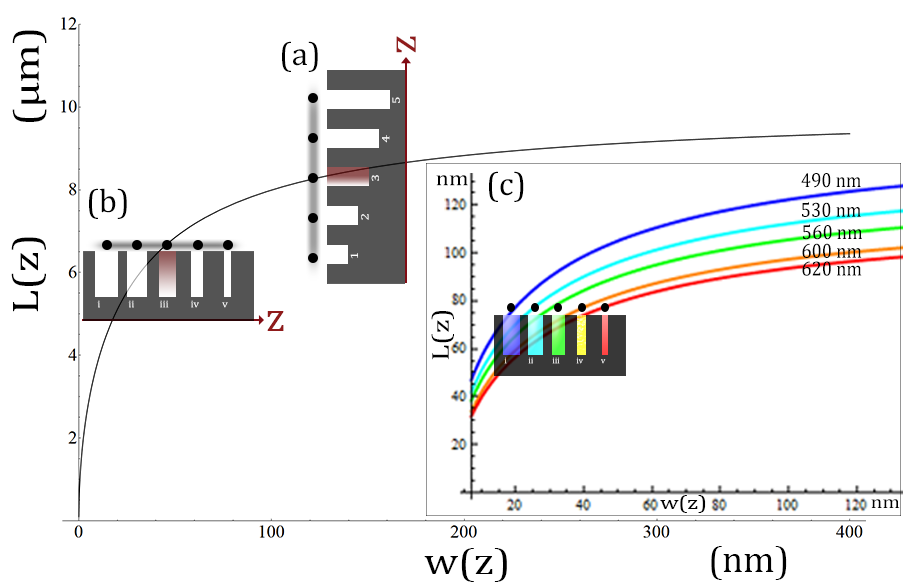}
\caption{\label{fig:structure_design} Plot of the fundamental cavity mode (n=0) is shown as a function $L$ and $w$ for a single frequency. Dotted lines (a) and (b) outline the discrete nature of the grating possessing linear gradients in $L$ and $w$, respectively, each point corresponding to a nano-groove of particular dimensions $(w_{ng},L_{ng})$. Inset (a): A grating with a linear gradient in depth variation corresponding to dotted line (a). Inset (b): A grating corresponding to dotted line (b), where the gradient is based on groove width. Inset (c) plots the resonant dispersion curve for several frequencies, each spectral component intersecting at different locations with the dotted line resulting in the formation of a rainbow trapping effect over the grating.}
\end{figure}

As previously mentioned, for most cavity resonators, $\lambda_{sp}$ is only a function of $L$, and in the case of gratings where $L$ changes along the $z$-direction (grooves gradually getting deeper or shallower) \cite{Bartoli} a spatial variation of cavity resonance is found with neither a variation in $\Gamma_{intra}$ nor a change in $n_{eff}$ (see inset (a) of Fig.~\ref{fig:structure_design}).
In the present Letter, the adiabatic variation of $\Gamma_{intra}$ (and consequently $w$) is introduced as a new dimension in the functional gradient space.

\begin{figure} [ht]
\includegraphics[height=55mm, width=87mm]{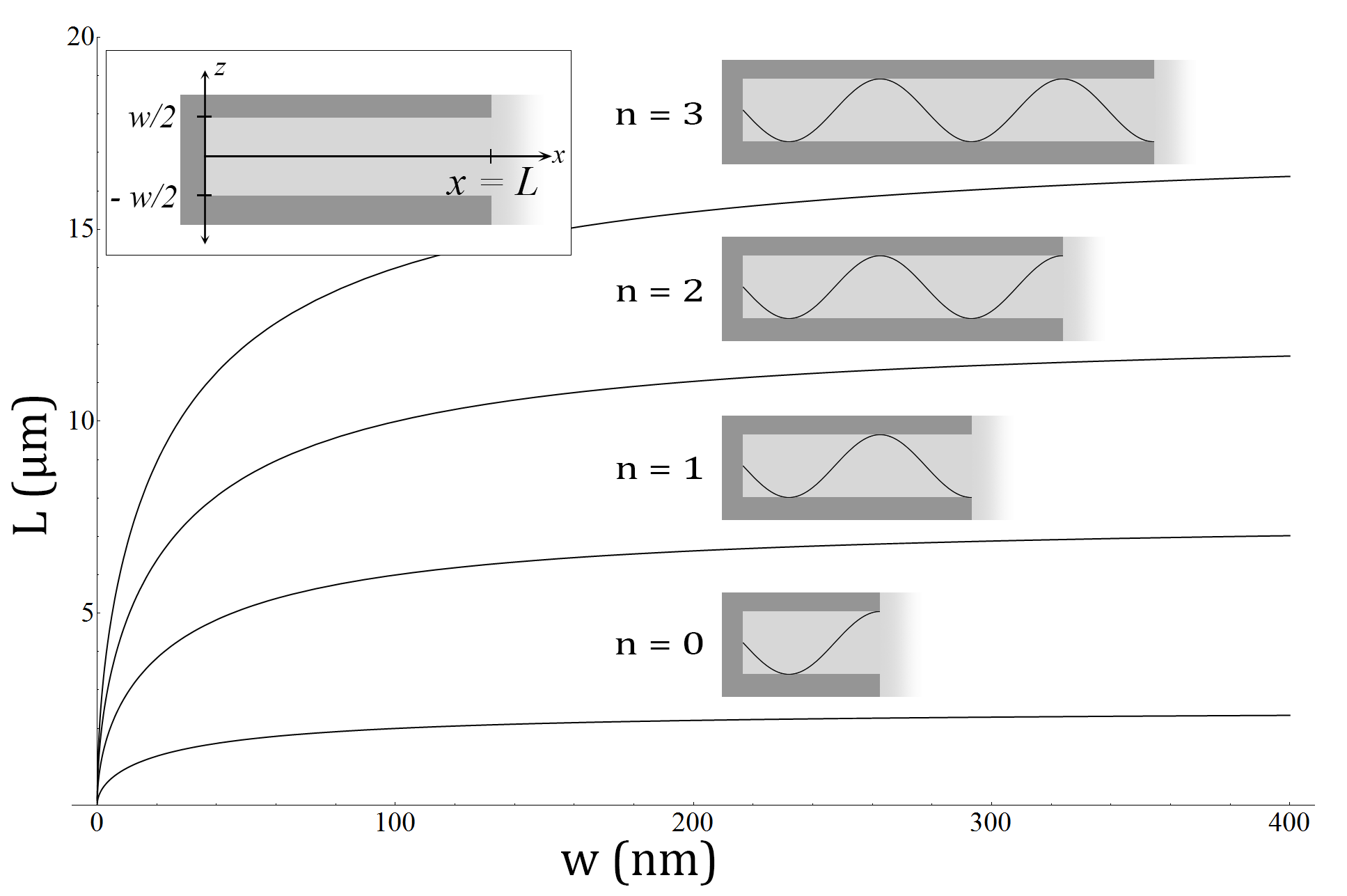}
\caption{\label{fig:multimodes} Discretized fundamental and higher order plasmonic modes given by Eqs. (\ref{eq:tan_dispersion}) and (\ref{eq:modes}). As $w$ exceeds $\sim150$ nm, modes approach Fabry-Perot cavity resonances (horizontal asymptotes). Conversely, reducing $w$ towards zero ramps up $\Gamma_{intra}$ and shortens the required $L$ of a resonator that manifests resonance at the same frequency. This is a known effect reported in \cite{Perchec}; in the electrostatic regime, shallower grooves can strongly absorb light, yet, deeper grooves are required if the intragroove coupling effects are not significant. It is also seen that compared to the fundamental mode, higher-$n$ modes are more sensitive to intragroove coupling effects.}
\end{figure}

We start by looking at the non-oscillatory $p$-polarized bound modes of an MIM waveguide which result in coupling of the localized modes in the core ($-\frac{w}{2}<z<\frac{w}{2}$ in the inset of Fig.~\ref{fig:multimodes}) (for details see \cite{plasmonic_book}), where the components of the magnetic and electric fields of the coupled-SPPs are given by:
\begin{eqnarray}
H_y = C e^{i\beta x}e^{k_1z}+D e^{i\beta x}e^{-k_1z},
\\
E_x = -iC \frac{1}{\omega \epsilon_0 \epsilon_1}k e^{i\beta x}e^{k_1z}+iD \frac{1}{\omega \epsilon_0 \epsilon_1}k e^{i\beta x}e^{-k_1z},
\\
E_z =C \frac{\beta}{\omega \epsilon_0 \epsilon_1}e^{i\beta x}e^{k_1z}+D \frac{\beta}{\omega \epsilon_0 \epsilon_1}e^{i\beta x}e^{-k_1z},
\label{eq:coupledCoreFields}
\end{eqnarray}
where $k_1$ and $k_2$ are the components of the $k$ vector perpendicular to the intragroove surface (i.e., along the $z$-direction) in the core and the metal, respectively, given by $k_1=\sqrt{\beta^2-\epsilon_1k_0^2}$, $k_2=\sqrt{\beta^2-\epsilon_2k_0^2}$, and where $k_0=\frac{\omega}{c}$, and $\omega$ is the frequency of the excitation.
By adding the two solutions in the core region of thickness $w$ and applying continuity conditions, the dispersion relation of the MIM waveguide is readily obtained. The odd modes are given by \cite{plasmonic_book}:
\begin{eqnarray}
\tanh\left (k_1 \frac{w}{2}\right)=-\frac{k_1\epsilon_1}{k_2\epsilon_2}.
\label{eq:tan_dispersion}
\end{eqnarray}

Fig.~\ref{fig:open_MIM} shows the surface plasmon wavelength $\lambda_{sp}$ = $\frac{2\pi}{\rm Re(\beta)}$ calculated from Eq.~(\ref{eq:tan_dispersion}) as a function of the core thickness $w$ \cite{Perchec}. 

For cavities with a single open end, applying the continuity conditions: $E_{z1}\big|_{x=0}$ + $E_{z2}\big|_{x=0}=0$ due to the perfect electric conductor at the bottom of the resonator, and $\frac{\partial(H_{y1}+H_{y2})}{\partial x}\big|_{x=L}=0$ due to the near unity reflection from the top of the resonator where $\beta\gg\sqrt{\epsilon}k_0$, yields the relation between the plasmonic wavelength, cavity length, and cavity modes:
\begin{eqnarray}
\left(\frac{1}{4}+\frac{n}{2}\right)\lambda_{sp}=L,
\label{eq:modes}
\end{eqnarray}
where $n$ is an integer denoting the mode order. Fig.~\ref{fig:multimodes} which depicts Eqs. (\ref{eq:tan_dispersion}) and (\ref{eq:modes}) together represents the dispersion diagram of a groove under resonance. In other words, it is the resonant dispersion relation of a plasmonic cavity as a function of its geometric dimensions $L$ and $w$. It is seen that for large values of $w$, $\lambda_{sp}$ approaches that of a classical Fabry-Perot type cavity whose resonant modes are independent of $w$ and determined simply by $L$.  However, the effect illustrated in Fig.~\ref{fig:open_MIM} for an unbounded MIM waveguide, and equally applicable to the quantized modes of a plasmonic groove shown in Fig.~\ref{fig:multimodes}, demonstrates width-dependence of resonant modes in extremely narrow plasmonic grooves due to the strong-coupling of SPPs,  in contrast to classical cavity resonators. This dependence is due to $\Gamma_{intra}$ which becomes noticeable when $w$ is around the metal skin depth, i.e. $w\lesssim$ 150 nm and quite significant when $w\lesssim$ 100 nm.

Using Eq. (\ref{eq:tan_dispersion}), $n_{eff}$ of each cavity is found to be: 
\begin{eqnarray}
 n_{eff}=\sqrt{\frac{\alpha^2\epsilon_1\epsilon_2^2-\epsilon_1^2\epsilon_2}{\alpha^2\epsilon_2^2-\epsilon_1^2}}
\label{eq:effective_index},
\end{eqnarray}
where $\alpha\equiv\tanh(\frac{k_1w}{2})$ for brevity. When $w\rightarrow0$ (bringing the metallic walls closer and shrinking the dielectric gap,) $n_{eff}\rightarrow \sqrt{\epsilon_2}$ and the effective index of the metal is recovered. It is worth mentioning that normally field penetration into metals is on the order of skin-depth which prevents electromagnetic waves from profoundly penetrating a medium of such index. However, through introduction of deep subwavelength grooves assisted by the strong-coupling of Type-1 SPPs and giving rise to a mode with no cutoff \cite{Prade}, the external fields are able to reach deep into the metal. Such a metasurface becomes particularly important for infrared wavelengths, where field penetration into the metals is nearly nonexistent and metals behave as nearly perfect conductors. This phenomenon is also fundamentally related to spoof SPPs where perforations or corrugations spoof surface plasmons with overall effects that are similar to real SPPs  \cite{Pendry_scienece}. At the other extreme, where $w\rightarrow\infty$ (separating one wall from the other, resulting in a single vertical wall) $n_{eff}=\sqrt{\frac{\epsilon_1 \epsilon_2}{\epsilon_1+\epsilon_2}}$ which is that of a flat metal-insulator interface. Engineering a grating with a tapered $n_{eff}$, through adiabatic tapering of $w$, results in a metasurface with a spatially varying effective index similar to a layered graded index material. Eq. (\ref{eq:effective_index}) shows that such a nano-structured surface defines an effective medium with an $n_{eff}\in [\sqrt{\frac{\epsilon_1 \epsilon_2}{\epsilon_1+\epsilon_2}}, \sqrt\epsilon_2).$ 

We now turn to examining gratings based on such tunable unit-resonators. In the regime of strong $\Gamma_{intra}$ (when $w\lesssim150$ nm) and under negligible $\Gamma_{inter}$ (when $d\gtrsim$ 100nm), we can segue from the response of a single groove to the response of the ensemble of grooves, i.e. the grating, using only the resonant dispersion relation of a single groove. 

It is worth mentioning that the nano-grooves need not be closed-ended cavities. For example, gratings based on open-ended waveguides result in a simple modification of Eq.~\ref{eq:modes} by changing $\frac{\lambda_{sp}}{2}$ to $\frac{\lambda_{sp}}{4}$, yielding an accordingly adjusted resonant dispersion curve.

Tailoring gratings using this intragroove coupling as a gradient parameter spans functionally graded structures based on $w$ as shown in inset (b) of Fig.~\ref{fig:structure_design}. The horizontal dotted line (b) in Fig.~\ref{fig:structure_design} represents the envelope function of a $\Gamma_{intra}$-based metasurface. Each dot corresponds to a groove whose $w$ is tapered as a function of the spatial variable shown in the corresponding inset (b) of Fig.~\ref{fig:structure_design}. A uniform depth grating, which is an ensemble of grooves of identical length $L$ with a gradually changing $w$ in the spatial direction $z$, guides light towards the location of a trap with a resonant mode that corresponds to the external wave's frequency. Each frequency component is then trapped at the respective spatial position, resulting in a rainbow-trapping shown in the inset (c) of the figure.

Full wave simulations carried out for the structures analyzed here show excellent agreement with this simple analytical model.
Since we are considering gratings with spatial gradients along the $z$-direction, it is useful to reintroduce $L$ and $w$ as functions of $z$ as is done in Fig.~\ref{fig:structure_design}. This is accomplished by scaling $L$ by a factor of $10^3$ to the unit distance along the $z$ axes, and $w$ by a  factor of unity. 
Then, the coordinates of a point on the main plot of Fig.~\ref{fig:structure_design} simultaneously convey the \textit{unit-cell-metric}, or the dimensions of a single groove such as ($L_3,w_3$) which correspond to point $3$ on the dotted line (b), as well as the \textit{ensemble-metric}, that is, the location ($z_3$) of that nano-groove within the grating shown in the inset figure (b). Additionally, the coincidence of the resonant dispersion curve with point 3 indicates the on-resonance condition of this nano-groove at the given frequency $f$, whereas all the other points (nano-grooves) belonging to the grating (grooves 1,2,4, and 5) are not at resonance at this frequency. This plot is identical to that of Fig.~\ref{fig:multimodes} except that the higher modes are omitted for clarity of presentation.
\begin{figure}[ht]
\includegraphics[height=53mm, width=87mm]{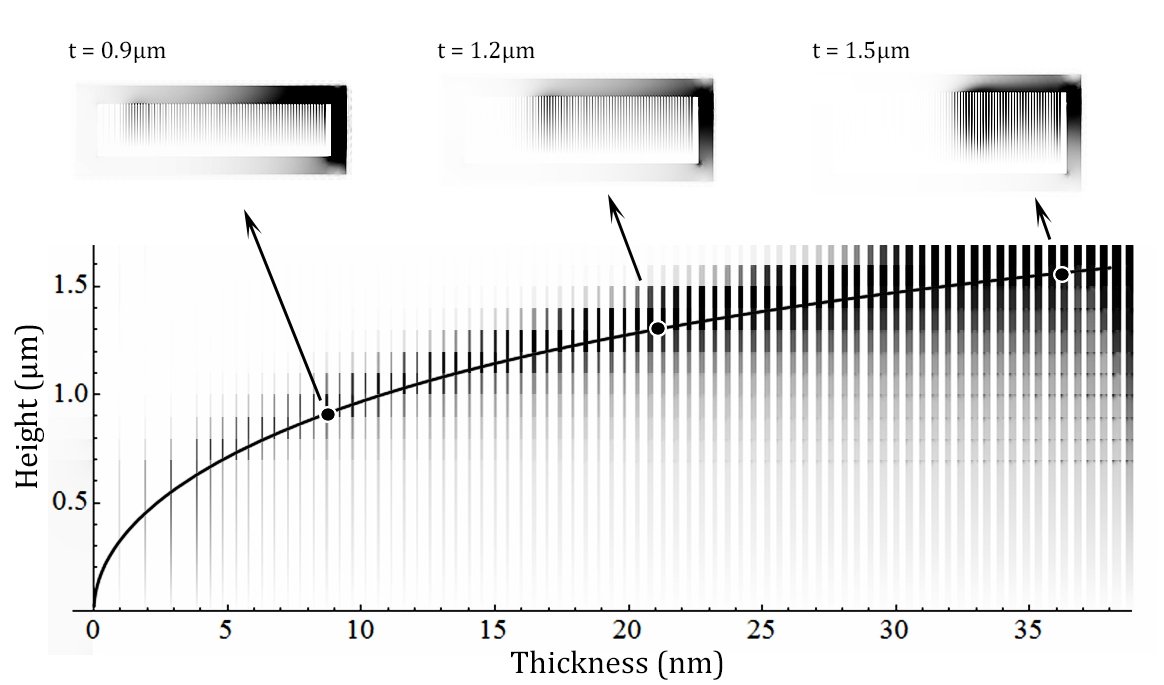}
\caption{\label{fig:super_graph} Several simulations are collated in successive frames, each showing the locus of light-trapping of a grating with the exact same width-gradient but a different depth. At each frame the groove depths are kept constant. All these structures  only utilize groove width (intergroove-coupling) as the gradient parameter. The curve on the projection screen behind all the frames shows the exact analytical solution of $\lambda_{sp}$ as a function of $L$ (groove depth) and $w$ (groove width) labeled as length and position axes respectively. When all these simulations are overlaid onto the analytical curve on the projection screen, the locations of light localization match with good accuracy. }
\end{figure}
With this convention, a unit distance increase or decrease along the vertical axis corresponds to a grating with deeper or shallower grooves by a unit distance (1 $\mu m$ here), and similarly a distance of 1 nm along the horizontal axis corresponds to varying the groove width by 1 nm. Therefore, successive points on the dotted lines shown in Fig.~\ref{fig:structure_design}, when taken individually, correspond to the responses of single grooves to monochromatic light of a given frequency, and when taken collectively, outline the discrete contour of a grating profile such as those depicted in the insets.
So long as the one-to-one correspondence between the groove geometry and its placement in the groove array (i.e., grating) is maintained, the intersection of the resonant dispersion plot in Fig.~\ref{fig:structure_design} with the dotted lines gives the location of light-trapping by that grating. This is because the resonant condition of Eq.~\ref{eq:modes} satisfied at all points along the plot, corresponds to grooves that trap radiation. It is worth mentioning that the dotted lines outlining the grating profiles, are only simple examples and arbitrary grating profiles can be analyzed just as easily. 

Fig.~\ref{fig:super_graph} shows overlays of numerical simulations of the electric field for ten different structures with the dark regions indicating high field localization at the location of the trap. Each simulation is performed with a structure that has constant $w(z)$. $L$ is successively increased in the subsequent simulation spanning the range of 0.5 $\mu$m to 1.5 $\mu$m. When the resulting curve is compared to the analytical model of Fig.~\ref{fig:structure_design}, the agreement is remarkably close.

Width-based gratings with uniform depth may prove to be useful tools for waveguiding and light trapping with applications in optoelectronics, infrared sensors, and energy devices. An advantageous consequence of gratings based on $\Gamma_{intra}$ compared to a depth gradient \cite{experimental_rainbow_bartoli} or a materials-based index gradient \cite{graded_fiber} is their facile and scalable fabrication which makes them practical candidates for a whole host of applications \cite{gradient_index_metamaterials}.	


\bibliographystyle{unsrt}

\end{document}